\documentclass[12pt]{article}

\batchmode

\newtheorem{theorem}{Theorem}
\newtheorem{lemma}{Lemma}

\newtheorem{definition}{Definition}

\newtheorem{corr}{Corollary}

\newcounter{aid}

\begin{document}

\title{ On Non-Markovian Performance Models }

\author{ \vspace*{1mm} Andr\'as  Farag\'o \\
Department of Computer  Science      \\
The  University   of  Texas   at  Dallas\\
   Richardson,  Texas \\
   {\tt farago@utdallas.edu} }

\date{}
\maketitle

\begin{abstract} {\bf\em We present an approach  that  can be useful when the
network  or  system  performance  is  described  by a model that is not
Markovian. Although most performance models are based on Markov chains
or Markov  processes, in  some cases  the  Markov 
property does not hold.
This  can  occur,  for  example,  when  the system exhibits long range
dependencies. For such situations, and other non-Markovian cases, our  method
can provide useful help. } \end{abstract}

\thispagestyle{empty}

\section{Introduction}

Network performance analysis is often  based on models that apply  the
mathematical  technique  of  Markov  chains  (or Markov processes when
continuous time is considered). Once we are able to set up a Markovian
model, we can investigate  both the stationary and  transient behavior
of the system, using well established methods. A classic example is the rich analysis of loss networks
in telecommunications, see Kelly \cite{kelly}.

In some   cases,  however, a  Markov model  cannot adequately
capture the behavior  of the system.  One possible reason  for this is
when the network behavior is inherently non-Markovian, possibly due to
long-range dependencies. We describe an approach that can help the analysis of
non-Markovian models, and bring back the possibility to apply results that are routinely used 
for Markov chains.

\section{General Setting: Discrete Stochastic Processes}

Let us consider stochastic processes with discrete time and
finite state space, without assuming that they are Markov chains. For
brevity, we call such a process a
{\em discrete stochastic process}.
We use the following notations:
\begin{itemize}

\item A discrete stochastic process: $\xi=(\xi_t,\; t=0,1,2,\ldots)$.

\item The state space of the process is assumed finite, and is denoted by $S$. Each
$\xi_t$ takes its values in $S$. The finiteness assumption can be relaxed, we just adopt it here for simplicity.

\item The probability distribution of $\xi_t$ is denoted by $\pi_t$,
which is identified with a vector in $[0,1]^{|S|}.$ (In matrix
expressions it will be regarded a row vector.)
We call these distributions the {\em one-dimensional distributions} of the
process.

\item We define  the {\em first-order transition probability matrix}
(or, simply, {\em transition probability matrix}) of
$\xi$    at     time    $t$     by
$$P_t=[p_t(a,b)]_{a,b\in     S}= $$
$$[\Pr(\xi_{t+1}=b\,|\,\xi_t=a)]_{a,b\in S}.$$
This is routinely used for Markov chains, but the conditional
probabilities can be defined for
any discrete stochastic process.  Note, however,  that if the process
is not a Markov chain, then, generally,
the value of $  \Pr(\xi_{t+1}=b\,|\,\xi_t=a) $ is not independent of
previous history, i.e., it may hold that 
\begin{equation}\label{nonmarkov}
\Pr(\xi_{t+1}=b\,|\,\xi_t=a)\neq
\Pr(\xi_{t+1}=b\,|\,\xi_t=a,\xi_{t-1}=a_{t-1},\ldots,\xi_0=a_0)
\end{equation}
which we refer to as {\em history dependence}.
\item If $P_t$ is  independent of $t$, then  we call the process  {\em
first-order homogeneous}. In this case all $P_t$ matrices can be
replaced
by      the       single matrix       $$P=[p(a,b)]_{a,b\in      S}= $$
$$[\Pr(\xi_{t+1}=b\,|\,\xi_t=a)]_{a,b\in S}.$$
Note that first-order homogeneity generally does not
imply the Markov property, so history dependence may still occur, i.e., we may still have (\ref{nonmarkov}).

\end{itemize}

\section{First-Order Equivalent Markov Chain}

Now we introduce a useful concept, called {\em First-Order Equivalent Markov Chain (1-EMC).}

\begin{definition}\label{def1} {\bf
(First-Order Equivalent Markov Chain (1-EMC))}
Let $\xi=(\xi_t,\; t=0,1,2,\ldots)$  be a discrete
stochastic process.
The {\em
First-Order Equivalent Markov Chain (1-EMC)}
of $\xi$  is defined as a Markov  chain
$\widetilde\xi=(\widetilde\xi_t,\;
t=0,1,2,\ldots)$ that is generated as follows: \begin{itemize}

\item Set $\widetilde\xi_0=\xi_0$.

\item Having obtained $\widetilde\xi_0,\ldots,\widetilde\xi_t$, the
value of
$\widetilde\xi_{t+1}$ is  drawn by
making an independent  random  transition from the
value of $\widetilde\xi_t$,
according  to  the  transition  probabilities  in $P_t$.

\end{itemize}

\end{definition}

We are also going to use the terminology that $\xi$ is the {\em parent
process}  of  $\widetilde\xi$.  It  is  clear from the definition that
$\widetilde\xi$ is indeed a Markov  chain, since it is generated  such
that whenever we are in a given state $a$ at time $t$, we move into a
state $b$ with probability $p_t(a,b)$ and this random choice is made, by definition,
independently of the previous history. (Note that even if the original process exhibits history dependence,
$p_t(a,b)$ is used as a {\em constant} probability for any given $t,a,b$.)
Consequently,
for every $a,b\in S$ and for every $t$
$$
\Pr(\widetilde\xi_{t+1} = b\,|\,\widetilde\xi_t=a) = p_t(a,b) =
$$
$$
\Pr(\widetilde\xi_{t+1}=b\,|\,\widetilde\xi_t=a,\widetilde\xi_{t-1}
=a_{t-1},\ldots,\widetilde\xi=a_0).
$$
Thus,
$\widetilde\xi$  has  the same first-order
transition probabilities as  the parent process  $\xi$, namely, $p_t(a,b)$. (On the  other
hand, generally this does not extend  to higher order
probability distributions if the parent
process is not a Markov chain.) Furthermore, it  is well known  from
the theory  of Markov  chains that  the initial  distribution and  the
(first-order) transition probabilities determine the chain uniquely,
so there is
no ambiguity when we talk about {\em the} 1-EMC of a discrete
stochastic process.  

\section{The Fundamental Property of the 1-EMC}

Let us now look at a key property of the First-Order Equivalent Markov Chain.

\begin{lemma}   \label{lemma1}    For   every    $t$   the    equality
$\widetilde\pi_t=\pi_t$ holds, where $ \widetilde\pi_t, \pi_t$ are the
one-dimensional  distributions  of  the  1-EMC and  the parent
process, respectively,  at time $t$. \end{lemma}

\noindent {\bf Proof.}
Assume there is an integer $\tau$ with
$\widetilde\pi_{\tau}\neq
\pi_{\tau}$ and choose $\tau$ such that it is the smallest such
integer. Since
$\widetilde\pi_0=\pi_0$ by definition, we have
$\tau\geq
1.$ Let us express $\pi_{\tau}(b)$ for an arbitrary $b\in S$. We can
write, using the law of total probability:
$$
\Pr(\xi_{\tau}=b) =
$$
$$
\sum_{a\in S} \Pr(\xi_{\tau}=b\,|\,\xi_{\tau-1}=a)
\Pr(\xi_{\tau-1}=a).
$$
With our notation this is
$$
\pi_{\tau}(b) =
 \sum_{a\in S} p_{\tau-1}(a,b) \pi_{\tau-1}(a)
$$
which in vector form gives
$$ \pi_\tau = \pi_{\tau-1} P_{\tau-1}.$$
By the choice of $\tau$ we have
$\widetilde\pi_{\tau-1}=\pi_{\tau-1}$, yielding
\begin{equation} \label{eq1}
 \pi_\tau = \widetilde\pi_{\tau-1} P_{\tau-1}.
\end{equation}
On the other hand, as
the first-order transition probabilities of $\xi$ and $\widetilde\xi$
are equal by the defining construction, we  obtain
that in the Markov chain $\widetilde\xi$
\begin{equation} \label{eq2}
 \widetilde\pi_\tau = \widetilde\pi_{\tau-1} P_{\tau-1}.
\end{equation}
holds. Comparing
(\ref{eq1}) and (\ref{eq2}) results in
$\widetilde\pi_\tau = \pi_\tau$, contradicting to the definition of
$\tau.$ Thus,
$\widetilde\pi_t=\pi_t$ must hold for every $t$.

\hfill $\spadesuit$

\section{Consequences}

\subsection{Trajectory Summation Formula}

An  important  consequence  of  Lemma~\ref{lemma1}  is that some basic
formulas that  are routinely  used for  Markov chains,  in fact remain
valid for {\em arbitrary} discrete stochastic processes.

\begin{corr}\label{corr1}
For every discrete stochastic process
\begin{equation}\label{corr1.1}
\pi_t=\pi_0 \prod_{i=0}^{t-1} P_i
\end{equation}
holds. The probability $\Pr(\xi_t=a)$ can be expressed as
\begin{equation}\label{corr1.1b} \Pr(\xi_t=a)\;=
\sum_{a_0,\ldots,a_t=a} \Pr(\xi_0=a_0)
p_0(a_0,a_1)\cdot\ldots\cdot p_{t-1}(a_{t-1},a_t)
\end{equation}
where the summation is taken over all trajectories
$a_0,a_1,\ldots,a_t$ with $a_t=a$.
Moreover, if the process is first-order homogeneous (but still not
necessarily Markov), then the above formulas simplify to
$$\pi_t=\pi_0 P^t$$ and
$$
\Pr(\xi_t=a)\;=
\sum_{a_0,\ldots,a_t=a}
\Pr(\xi_0=a_0) p(a_0,a_1)\cdot\ldots\cdot p(a_{t-1},a_t).
$$
\end{corr}

\noindent  {\bf  Proof.}  For  the  1-EMC  of  $\xi$  the  relationship
$\widetilde\pi_t=\widetilde\pi_0 \prod_{i=0}^{t-1} P_i$ holds, being a
Markov chain. By  Lemma~\ref{lemma1} we have  $\widetilde\pi_t=\pi_t$,
implying (\ref{corr1.1}). If we write  down the details of the  matrix
product in (\ref{corr1.1}), we get precisely (\ref{corr1.1b}). If $\xi$
is  first-order  homogeneous,  then  $P_0=P_1=\ldots=P_t$  holds, too,
yielding the second pair of formulas.

\hfill $\spadesuit$

\medskip\medskip
Note that if $\xi$ is a Markov chain (possibly not time-homogeneous),
then the probability that
we reach $a_t$ via a given trajectory $a_0,a_1,\ldots,a_t$ is precisely the
product
\begin{equation}\label{prod}
\Pr(\xi_0=a_0) p_0(a_0,a_1)\cdot\ldots\cdot
p_{t-1}(a_{t-1},a_t)
\end{equation}
due to the Markov property. Since reaching $a_t$ via different
trajectories are exclusive events and they represent all
possibilites, therefore,
summing up for all such possible products naturally gives the formula
\begin{equation}\label{corr1.2}
\Pr(\xi_t=a)\;=
\sum_{a_0,\ldots,a_t=a} \Pr(\xi_0=a_0)
p_0(a_0,a_1)\cdot\ldots\cdot p_{t-1}(a_{t-1},a_t)
\end{equation}
for Markov chains. On the other hand, if $\xi$ is {\em
not} a Markov
chain, then the probability of traversing a given trajectory
$a_0,\ldots,a_t$ may not
be equal to (\ref{prod}) because of the effect of history dependence.
 Nevertheless, the trajectory {\em summation} formula
(\ref{corr1.2}) still remains valid, even though the individual
summands may not be equal to the individual probabilities of the corresponding trajectories.

\subsection{Stationary Distribution}

Via  the  1-EMC,  we  can  directly  carry over a number of fundamental
concepts  and  results  from  Markov  chain  theory  to a more general
setting. Let us look at the stationary distribution.

\begin{definition}\label{defcommon}
Let $\xi=(\xi_t,\; t=0,1,2,\ldots)$  be a discrete
stochastic process with state space $S$. Assume that $\xi$ is
first-order homogeneous (but may not be Markov) and let its
first-order transition probability matrix be $P$.
\begin{itemize}

\item
A probability distribution $\pi$ on $S$ is called a
{\em stationary distribution} of $\xi$ if $\pi=\pi P$ holds.

\item A process is called {\em ergodic} if it has a stationary distribution $\pi$, and the one-dimensional distribution $\pi_t$ 
satisfies $\lim_{t\rightarrow\infty} \pi_t=\pi$.

\item The process is called {\em irreducible} if there exists a
positive integer $k$ with $P^k>0,$  that is, every entry of the matrix $P^k$ is positive.

\item The process is called {\em aperiodic} if for every $a\in S$
$${\rm gcd}\{m:\; p^{(m)}(a,a)>0\}=1$$ holds, where the
$p^{(m)}(.,.)$ are the entries of $P^m$, and {\rm gcd} means greatest common divisor.
\end{itemize}
\end{definition}

The concepts of Definition~\ref{defcommon} are routinely used for Markov chains, but they do not actually require the Markov property, 
so they can be extended to arbitrary first-order homogeneous discrete stochastic processes.

Now we can state how the fundamental features of these concepts  carry
over from Markov chains to arbitrary first-order homogeneous  discrete
stochastic processes.

\begin{theorem} \label{thm1}
Let $\xi=(\xi_t,\;  t=0,1,2,\ldots)$ be a  first-order
homogeneous discrete stochastic process. Assume that $\xi$ is
irreducible and aperiodic (but may not be Markov). Then the
following hold:
\begin{itemize}

\item The process is ergodic, i.e., it has a unique stationary distribution $\pi$, and 
$\lim_{t\rightarrow\infty}\pi_t=\pi$ holds.

\item  The  1-EMC  of  $\xi$  also has  a  unique  stationary   distribution
$\widetilde\pi$. Moreover, $\widetilde\pi=\pi$, and the 1-EMC is an ergodic Markov chain.

\item  The rate of convergence to stationarity in $\xi$ is the  same
as  in the  1-EMC. In particular, $\pi_t-\pi=\widetilde\pi_t-\widetilde\pi$ holds for every $t$.

\end{itemize}
\end{theorem}

\noindent  {\bf   Proof.}  By   the  definition   of  the   1-EMC,  the
(first-order) transition probability matrix is the same for $\xi$  and
$\widetilde\xi$.
  By  Lemma~\ref{lemma1}  we  have  $\widetilde\pi_t=\pi_t$
for
every $t.$ The rest follows  directly from the well known  fundamental
results of Markov chain theory on the stationary distribution and ergodicity, see, e.g., \cite{kemeny,kijima,norris}.

\hfill $\spadesuit$

\subsection{Censored Discrete Stochastic Processes}

\noindent
The concept of {\em censoring} is well known in the Markov chain setting; it means that we only
observe the chain when it is in a given subset, the rest is ``censored out." Below we explain it in some more details.

Let $X=(X_t, \, t=0,1,2,\ldots)$  be a Markov chain, and let $A\subseteq S$ be a nonempty 
subset of states.  
The {\em censored Markov chain} (with respect to $A$), which is also referred to as the {\em chain watched only on} $A$,
is defined by keeping only those members of the sequence that fall in $A$ (the $A$-hits), the rest are censored out.
That is,  if $\tau_0<\tau_1<\tau_2<\ldots$ are the (random) times of all the $A$-hits, then the censored Markov chain is represented by the sequence
$$Y=(Y_t, \, t=0,1,2,\ldots)=(X_{\tau_t},\,t=0,1,2,\ldots).$$ Interestingly, the censored sequence $Y$ remains a Markov chain, with state space
$A$. The (nontrivial) fact that $Y$ satisfies the Markov property is a consequence of the Strong Markov Property 
(for a derivation see, e.g., Norris \cite{norris}).
There is also an expression for the transition probabilities of
the censored chain, but we are not going to need it.

Another feature of the censored Markov chain is that if  the original chain is ergodic with stationary distribution $\pi$,
then the censored chain remains ergodic with stationary distribution 
\begin{eqnarray} \label{pia}
\pi_A(x) = \left\{
    \begin{array}{ccl}
      \pi(x)/\pi(A)  & \;\;\mbox{ if }\;\;\;  x\in A\\
      0 & \;\;\mbox{ if }\;\;\;   \, x\notin A.
    \end{array} \right.
\end{eqnarray}
 This fact follows from the Ergodic Theorem, see, e.g.,  Aldous and Fill \cite{aldous}, Section 2.7.1. Note that $\pi_A$ is just the original stationary stationary distribution conditioned on the set $A$. A simple, but useful consequence is stated in the next lemma.

\begin{lemma} \label{censor}
Let   $X=(X_t, \, t=0,1,2,\ldots)$ be   an ergodic    Markov  chain, with stationary distribution $\pi$, and $A\subseteq S$, $A\neq\emptyset$.
Let the distribution of $X_0$ be $\pi_A$, given in {\rm (\ref{pia})}. Then each $A$-hit has the same distribution $\pi_A$.
\end{lemma}

\noindent
{\bf Proof.} Let $\tau_0<\tau_1<\tau_2<\ldots$ be the random times 
of all the $A$-hits. As discussed above,  $Y=(X_{\tau_t},\,t=0,1,2,\ldots)$ remains an ergodic Markov chain with stationary distribution $\pi_A$.
Since we start $X$ from the initial distribution $\pi_A$, therefore, $\tau_0=0$. It means, the censored chain $Y=(X_{\tau_t},\,t=0,1,2,\ldots)$, 
which is obtained by keeping only the $A$-hits from the original chain, starts from its stationary distribution, so it must also remain in the same distribution. Consequently, each $A$-hit $X_{\tau_t}$ is distributed by $\pi_A$. (This holds even if the original chain $X$ may not be in its stationary distribution $\pi$ at any given time.)

\hfill $\spadesuit$

\medskip
The above results carry over to discrete stochastic processes, as well, via the 1-EMC. Specifically, we have the following 
theorem:

\begin{theorem} 
Let $\xi=(\xi_t,\;  t=0,1,2,\ldots)$ be a  first-order
homogeneous discrete stochastic process. Assume that $\xi$ is
irreducible and aperiodic, with stationary distribution $\pi$ (but it may not be Markov). Further, let $A\subseteq S$, $A\neq\emptyset$, and 
let the distribution of $\xi_0$ be $\pi_A$, given by the formula {\rm (\ref{pia})}.   Then each $A$-hit of $\xi$ has the same distribution $\pi_A$.
\end{theorem}

\noindent
{\bf Proof.} By Theorem~\ref{thm1}, $\xi$ has a  stationary distribution $\pi$. Let $X=(X_t, \, t=0,1,2,\ldots)$ be the 1-EMC of $\xi$. 
Again by Theorem~\ref{thm1}, $X$ is an ergodic Markov chain with stationary distribution $\pi$. Then, by Lemma~\ref{censor}, we have that each
$A$-hit of $X$ has the same distribution $\pi_A$, given by the formula {\rm (\ref{pia})}. Since by Lemma~\ref{lemma1} the 
one-dimensional  distributions  of  the  1-EMC and  the parent process coincide, therefore, 
each $A$-hit of $\xi$ has the same distribution $\pi_A$, as well.

\hfill $\spadesuit$

\section{Conclusion}

The presented results provide a method to handle non-Markovian models.
If  we  are  able  to  deduce  or  measure  the first-order transition
probabilities of the system, then the stationary distribution and also
the speed of convergence  to stationarity (transient analysis)  can be
obtained from the analysis of  the 1-EMC, utilizing the fact  that its
1-dimensional  distributions  coincide  with  that  of  the   original
process. In other words, we can reduce the analysis of a non-Markovian
system to a Markov chain, carrying over a number of results from Markov chain theory.

\end{document}